\documentclass[twocolumn]{revtex4}

\usepackage{amsfonts,amsmath,amssymb,amsthm}
\usepackage{graphicx}

\usepackage{graphicx}
\usepackage{dcolumn}
\usepackage{bm}

\newcommand{\dd}{\mathrm{d}}
\newcommand{\beq}{\begin{equation}}
\newcommand{\eeq}{\end{equation}}
\newcommand{\beqr}{\begin{eqnarray}}
\newcommand{\eeqr}{\end{eqnarray}}
\newcommand{\beqrn}{\begin{eqnarray*}}
\newcommand{\eeqrn}{\end{eqnarray*}}
\newcommand{\beqn}{\begin{equation*}}
\newcommand{\eeqn}{\end{equation*}}
\newcommand{\bei}{\begin{itemize}}
\newcommand{\beii}{\begin{itemize} \item}
\newcommand{\eei}{\end{itemize}}
\newcommand{\ben}{\begin{enumerate}}
\newcommand{\een}{\end{enumerate}}
\newcommand{\bes}{\begin{small}}
\newcommand{\ees}{\end{small}}
\newcommand{\bec}{\begin{center}}
\newcommand{\eec}{\end{center}}

\usepackage[usenames,dvipsnames]{xcolor}

\begin{document}


\title{A simple mechanism for higher-order correlations\\ in integrate-and-fire neurons}

\author{David A. Leen}

 
\author{Eric Shea-Brown}%
\email{etsb@washington.edu}
\affiliation{%
Department of Applied Mathematics, University of Washington.
}

\date{\today}

\begin{abstract}


The collective dynamics of neural populations are often characterized in terms of {\it correlations} in the spike activity of different neurons.  Open questions surround the basic nature of these correlations.  In particular, what leads to higher-order correlations -- correlations in the population activity that extend beyond those expected from cell pairs?  Here, we examine this question for a simple, but ubiquitous, circuit feature:  common fluctuating input arriving to spiking neurons of integrate-and-fire type.  We show that leads to strong higher-order correlations, as for earlier work with discrete threshold crossing models.  Moreover, we find that the same is true for another widely used, doubly-stochastic model of neural spiking, the  linear-nonlinear cascade.  We explain the surprisingly strong connection between the collective dynamics produced by these models, and conclude that higher-order correlations are both broadly expected and possible to capture with surprising accuracy by simplified (and tractable) descriptions of neural spiking.

 \end{abstract}

\pacs{87.19.lj}
\maketitle


Interest in the collective dynamics of neural populations is exploding, as new recording technologies yield views into neural activity on larger and larger scales, and new statistical  analyses yield potential consequences for the neural code~\cite{Shlens2009,Ganmor:2011ct,BrownKM04,Pillowetal08,ZylberbergSB12}.  A fundamental question that arises as we seek to quantify these population dynamics is the statistical {\it order} of interactions among spiking activity in different neurons.  That is, can the co-dependence of spike events in a set of neurons be described by an (overlapping) set of correlations among pairs of neurons, or are there irreducible higher-order dependencies as well?  Recent studies show that purely pairwise statistical models are successful in capturing the spike outputs of neural populations under some stimulus conditions~\cite{Sch:2006,Shlens:2006,Shlens2009}, but that different populations or stimuli can produce beyond-pairwise interactions~\cite{Ohiorhenuan:2010bu,Ganmor:2011ct,Montanietal09,Yu:2011ii}.


 
Despite these rich empirical findings, we are only beginning to understand what {\it dynamical} features of neural circuits determine whether or not they will produce substantial higher-order {\it statistical} interactions.  Recent work has suggested that one of these mechanisms is common -- or {correlated} -- input fluctuations arriving simultaneously at multiple neurons~\cite{macke11,Amarietal03,BarreiroGRS10,Sch+03,Koster:2013}; importantly, this is a feature that occurs many  neural circuits found in biology~\cite{Sha+98,Bin+01,tr08}.  In particular,~\cite{macke11,Amarietal03} showed that common, gaussian input fluctuations, when ``dichotomized" so that inputs over a given threshold produce spikes, give rise to strong beyond-pairwise correlations in the spike output of large populations of cells.  This is an interesting result, as a step function thresholding mechanism produces higher-order correlations in spike outputs starting with purely pairwise (gaussian) inputs.  

A natural question is whether more realistic, dynamical mechanisms of spike generation -- beyond ``static" step function transformations -- will also serve to produce strong higher-order correlations based on common input processes.  In this paper, we show that the answer is yes, and connect several widely-used models of neural spiking to explain why.  We begin with a model of integrate-and-fire type.

\medskip

\paragraph*{An exponential integrate-and-fire population with common inputs.}

\begin{figure}[b!]
\includegraphics{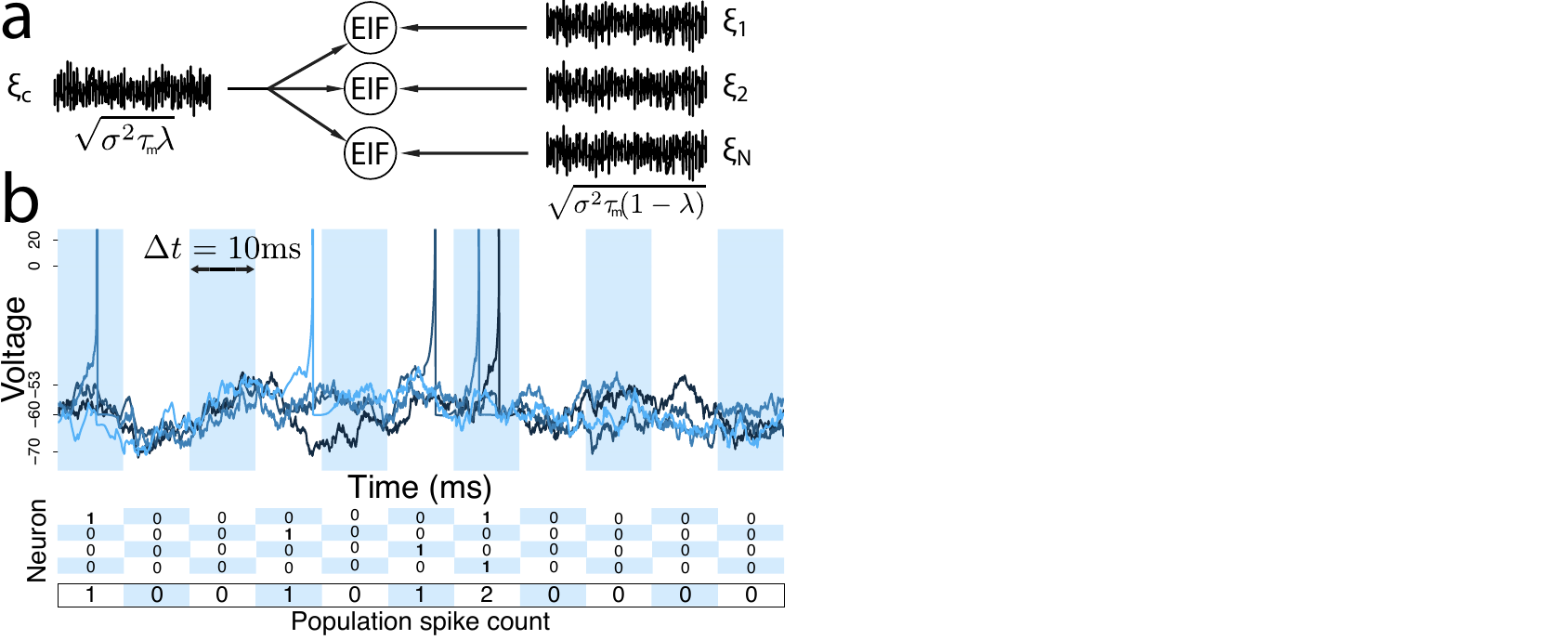}
\caption{\label{fig:schematic} (a) EIF neurons receiving common $\xi_c$ and independent inputs $\xi_i$. The voltages of the neurons evolve according to Eqn.~\eqref{eifsde}. Parameters: $\tau_m = 5$ms, $\Delta_T = 3$mV, $V_T = 20$mV, $V_S = -53$mV, $V_R = -60$mV, $\tau_{\text{ref}} = 3$ms.  We tune the noise amplitude so when the DC component of the input is $\gamma = -60$mV, the neurons fire at $10$Hz; this yields $\sigma = 6.23$mV. (b) Cartoon of the binning process: spikes recorded from each of the EIF neurons in a bin contribute towards the population spike count. More than one spike occurring {\em from the same neuron} within a single bin is treated as a single event. This happens less than $0.4\%$ of the time in our numerical simulations with $\mu=0.1$ and $\rho=0.1$.}
\end{figure}

Fig.~\ref{fig:schematic} shows a ubiquitous situation in neural circuitry:  a group of cells receiving common input.  We model this in a homogeneous population of $N$ exponential integrate-and-fire (EIF) neurons~ \cite{FourcaudTrocme03,Day+01}.  Each cell's membrane voltage $V_i$ evolves according to: 
\begin{align}
\label{eifsde}
\tau_m V_i^\prime &= -V_i +\psi(V_i)+I_i(t),\\
I_i(t) &= \gamma+\sqrt{\sigma^2\tau_m}\big[\sqrt{1-\lambda}\xi_i(t)+\sqrt{\lambda}\xi_c(t)\big] \nonumber,
\end{align}
where $\psi(V_i) =\Delta_T \exp{\left((V_i - V_S)/\Delta_T\right)}$.  Here, $\tau_m$ is the membrane time constant, $\Delta_T$ gives the slope of the action potential initiation, and $V_S$ the ``soft'' threshold.  When voltages cross a ``hard" threshold $V_T = 20$mV, they are said to fire a spike, and are reset to the value $V_R = -60$mV and held at that voltage for a refractory period $\tau_{\text{ref}} = 3$ms.  See the caption of Fig.~\ref{fig:schematic} for further parameter values, which drive the cell to fire with the typical irregular, poisson-like statistics~\cite{Sof+93}.

Each cell's input current $I_i(t)$ has a constant (DC) level $\gamma$, and a white noise term with amplitude $\sigma$.    The noise term has two components.  The first is the common input $\xi_c(t)$ which is shared among all neurons.  The second is an independent white noise $\xi_i(t)$; the relative amplitudes are scaled so that the inputs to different cells are correlated with (Pearson's) correlation coefficient $\gamma$ (as in, e.g.,~\cite{Lin+05,delaRocha:2007,SheaBrown:2008}, cf.~\cite{Bin+01}).

We quantify the population output by binning spikes with temporal resolution $T_\text{bin} = 10$ms (see Fig.~\ref{fig:schematic}). (On rare occasions ($<0.4\%$ of bins, see Fig.~\ref{fig:schematic} caption) multiple spikes from the same neuron can occur in the same bin. These are considered as a single spike.)  The spike {\it firing rate} is quantified by $\mu$, the probability of a spike occurring in a bin for a given neuron.  {\it Pairwise correlation} in the simultaneous spiking of neurons $\i,j$ is quantified by the correlation coefficient $\rho=Cov(n_i,n_j)/Var$, where $n_i$, $n_j$ are the $\{0,1\}$ spike events for the cells and $Var$ is their (identical) variance $\mu (1-\mu)$.


\medskip

\paragraph*{Emergence of strong higher-order correlations in EIF populations.} 

\begin{figure}[b]
\includegraphics{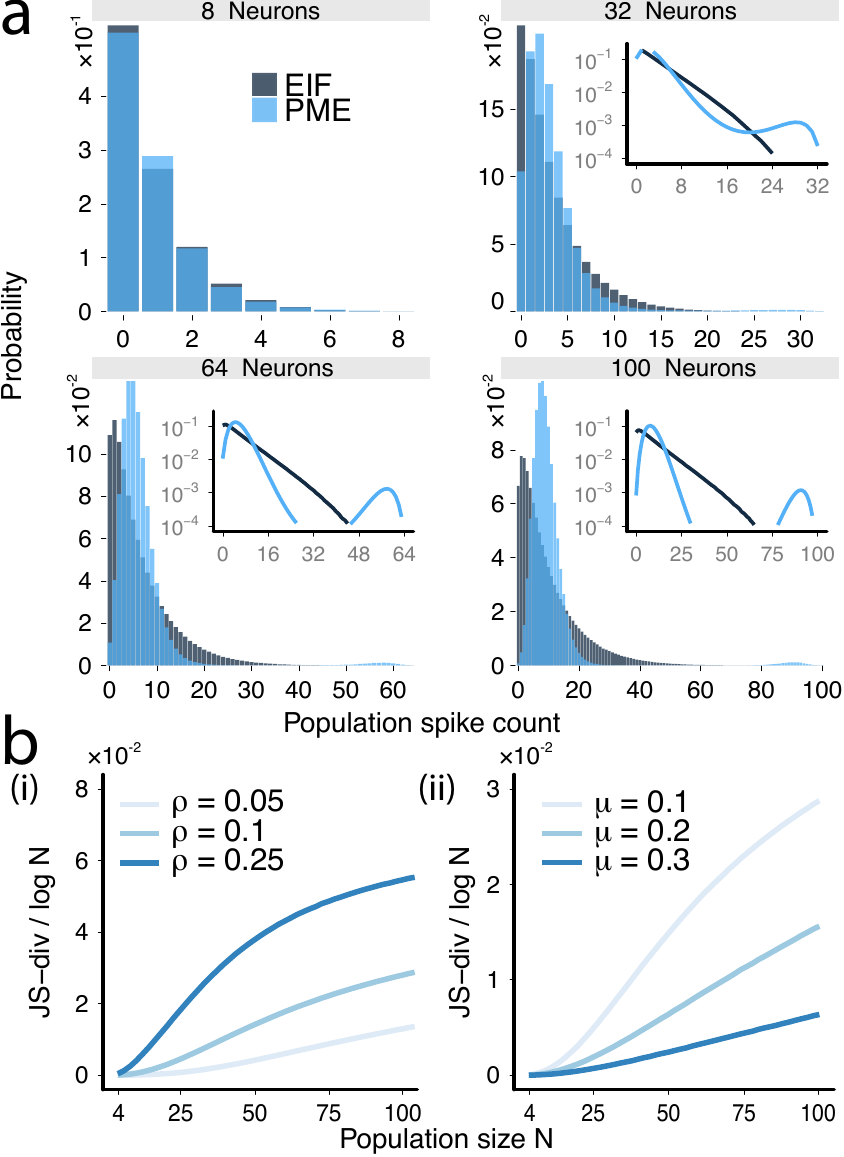}
\caption{\label{fig:eifising} (a) Population spike-count distributions $P_{\text{EIF}}(k)$ for the EIF and $P_{\text{PME}}(k)$ for its second-order approximation for $8,~32,~64,~\text{and } 100$ neurons, for $\mu = 0.1$ and $\rho = 0.1$. The distributions are similar for smaller populations but not for larger populations. Inset: the same distributions on a log-linear scale. (b) The Jensen-Shannon (JS) divergence between the EIF and the pairwise maximum entropy (PME) model. We normalize by $\log(N)$, the natural growth rate of the JS-divergence. (Left) JS-divergence for a constant value of $\mu = 0.1$ and increasing values of correlation $\rho$, and (Right) for constant value of $\rho = 0.1$ and increasing values of firing rate $\mu$, vs. population size. The JS-divergence grows with increasing $\rho$ and decreasing $\mu$.}
\end{figure}

Beyond these statistics of single cells and cell pairs, we describe multineuron activity via the distribution of population spike counts -- i.e.,~the probability $P_{\text{EIF}}(k)$ that $k$ out of the $N$ cells fire simultaneously (as in~, e.g., \cite{macke11,BarreiroGRS10,Montanietal09,Amarietal03}).  Fig.~\ref{fig:eifising}(a) illustrates these distributions.  The question we ask is:  
Do beyond-pairwise correlations play an important role in determining the population-wide spike count distribution? 

To answer this, we compare the population spike-count distribution $P_{\text{EIF}}(k)$ from the EIF model vs. its second-order approximation via a pairwise maximum entropy (PME) model with matched mean $\mu$ and pairwise correlation $\rho$:  $P_{\text{PME}}(k)= Z^{-1} \binom{N}{k}\exp{(\alpha k + \beta k^2)}$ with parameters fit numerically (for details see~\cite{macke11}).  Fig.~\ref{fig:eifising}(a) demonstrates that, for small populations, the PME and EIF distributions are similar.  However, for populations larger than about $N=30$ neurons, strong differences emerge.  This demonstrates that the EIF model produces higher-order correlations that strongly impact the structure of population firing.

We quantify the discrepancy between $P_{\text{PME}}(k)$ and $P_{\text{EIF}}(k)$ via the (normalized) JS-divergence.  See Fig.~\ref{fig:eifising}(b), which shows very similar results for the EIF system as those found for a thresholding model in~\cite{macke11} (see below).  In particular, the EIF model produces strong beyond-pairwise correlations at a wide range of correlations $\rho$ and mean firing rates $\mu$. Additionally, as in~\cite{macke11} (cf.~\cite{RoudiNL09}), the Jensen-Shannon divergence grows with increasing population size $N$. Moreover, the divergence increases with increasing pairwise correlation and decreasing mean firing rate.


\begin{figure}[t!]
\includegraphics{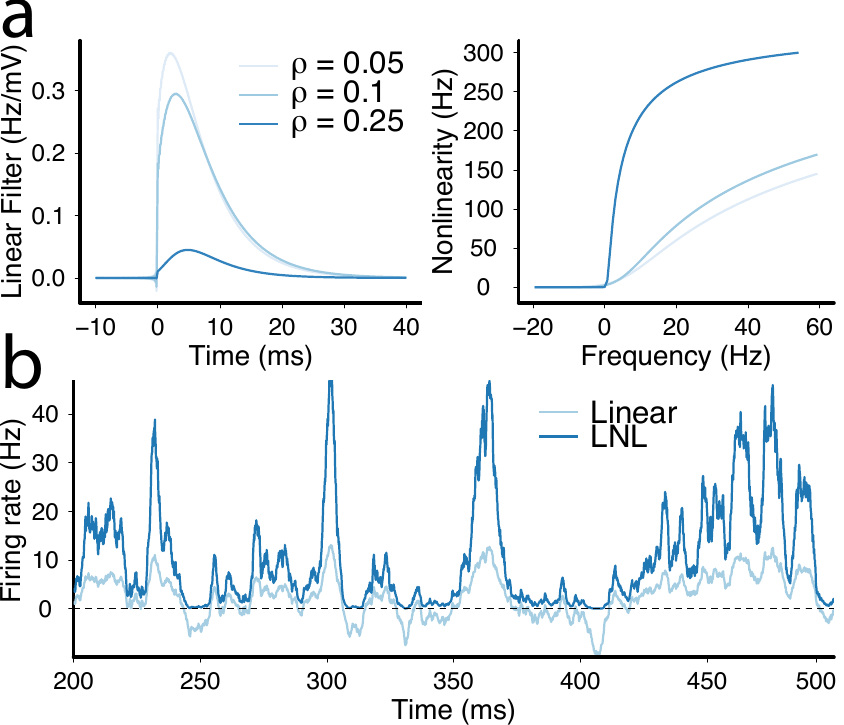}
\caption{\label{fig:linearfilter} (a) The linear filter $A(t)$ and static non-linearity $F$ computed for inputs that yield several values of the spike correlation coefficient $\rho$. The filter receives a noise amplitude of $\sigma\sqrt{1-\lambda}$. The static-nonlinearity receives a noise amplitude of $\sigma$. (b) The static non-linearity applied to the linear estimate of the firing rate, for $\mu = 0.1$, $\rho = 0.1$. The non-linearity increases the firing rate magnitude and rectifies negative firing rates.}
\end{figure}

\medskip 

\begin{figure}[t]
\includegraphics{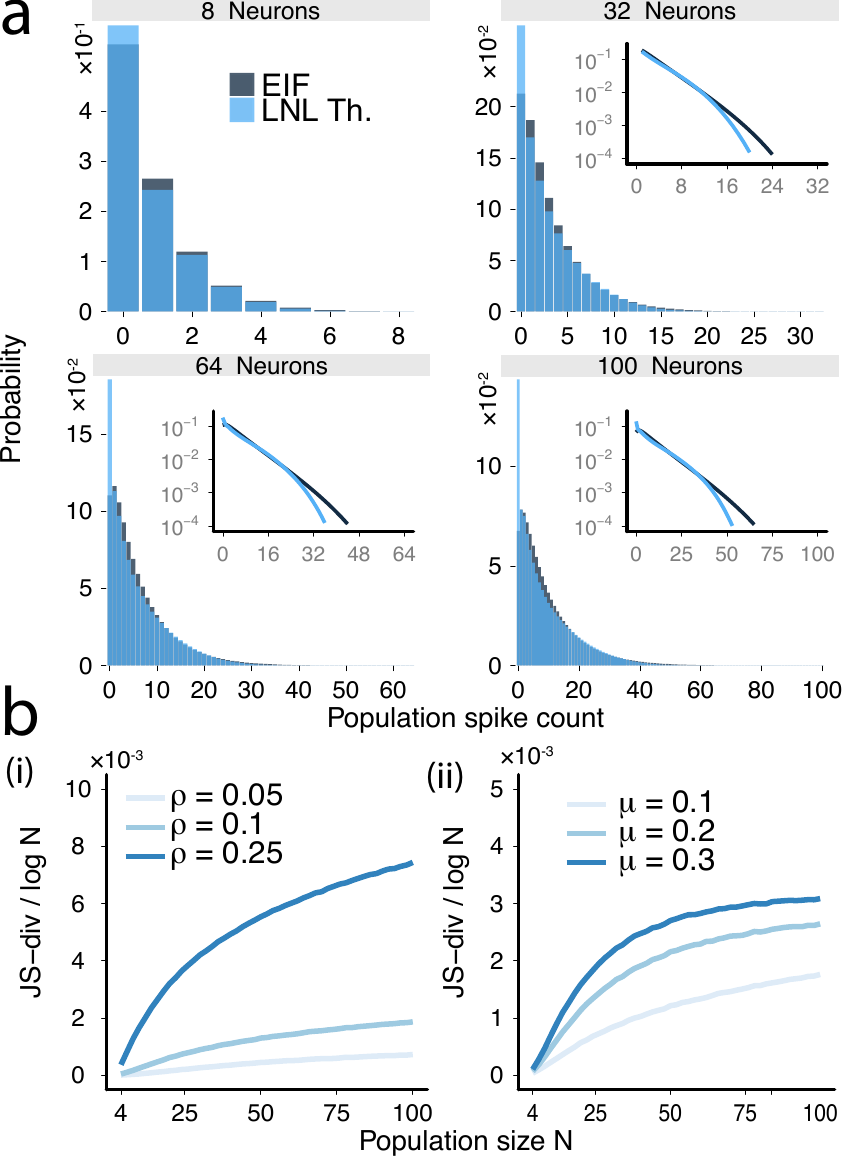}
\caption{\label{fig:eiffilter} (a) Population spike-count distributions $P_{\text{EIF}}(k)$ for the EIF model and $P_{\text{LNL}}(k)$ for the linear-nonlinear cascade approximation for $8,~32,~64,~\text{and } 100$ neurons for $\mu = 0.1$ and $\rho = 0.1$. While the trends are very similar overall, the LNL model greatly overestimates the zero population spike count probabilities and underestimates the tails. Inset: the same distributions on a log-linear scale. (b) The JS-divergence between the EIF and LNL is an order of magnitude smaller than PME. (Also, the order of the mean firing rates is reversed when compared to the PME as the LNL cascade gives a better approximation at higher firing rates.)}
\end{figure}

\paragraph*{A linear-nonlinear cascade model that approximates EIF spike activity and produces higher-order correlations.}  We next study the impact of common input on higher-order correlations in a widely-used point process model of neural spiking.  This is the linear-nonlinear cascade model, where each neuron fires as a (doubly-stochastic) inhomogeneous Poisson process.   We use a specific linear-nonlinear cascade model that is fit to EIF dynamics.  This both establishes that the common input mechanism is sufficient to drive higher-order correlations in the cascade model, and develops a semi-analytic theory for the population statistics in the EIF system.

In the linear-nonlinear cascade, each neuron fires as an inhomogeneous Poisson process with rate given by convolving a temporal filter $A(t)$ with an input signal $c(t)$ and then applying a time independent nonlinear function \cite{Ostojic:2011} $F$:   
\begin{equation}
r(t) = F\left(A * c(t) \right). \nonumber
\end{equation}
The signal for each cell is the common input: $c(t) = \sqrt{\sigma^2 \tau \lambda}~\xi_c(t)$. The filter $A(t)$ is computed as the linear response of the firing rate to a weak input signal, via an expansion of the Fokker Planck equation for Eqn.~\eqref{eifsde} around the equilibrium obtained with ``background'' current $\gamma+ \sqrt{\sigma^2 \tau (1-\lambda)}~\xi(t)$ . This calculation follows exactly the methods described in \cite{Richardson:2007}. For the static nonlinearity, we follow \cite{Ostojic:2011} and take $F(x) = \Phi\left( \gamma + \frac{x}{\Phi^\prime(\gamma)}\right)$, where $\Phi\left( \gamma \right)$ is the equilibrium firing rate obtained at the background currents described above.  This choice, in particular, ensures that we recover the linear approximation $r(t) = A * c(t)$ for weak input signals. The linear filter must be approximated numerically hence the semi-analytic nature of our model. The numerical approximations for the filter, nonlinearity, and resulting firing rate are shown in Fig.~\ref{fig:linearfilter}.

\begin{figure*}[t]
\includegraphics{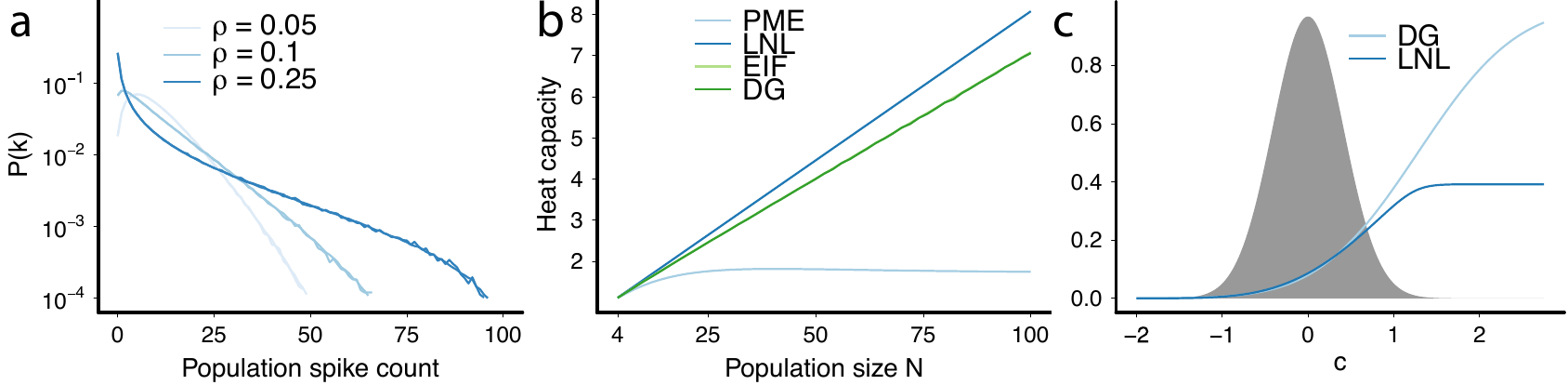}
\caption{\label{fig:lnldgcomp} 
(a) The Dichotomized Gaussian (DG) model gives an excellent description of the EIF population spike count probability distributions across a range of correlation coefficient values. The two models are plotted on top of one another and appear as a single curve for each value of $\rho$. (b) The heat capacity increases linearly for the LNL-cascade, the EIF and the DG. The heat capacity of the LNL cascade increases at a slightly greater rate than the EIF/DG which overlap. The Ising model saturates at a population of approximately $N=30$ neurons. (c) Comparing the $L(c)$ vs. the $\tilde L( f(c))$ functions for the DG and LNL models, respectively. The two functions largely agree over about 2 standard deviations of the Gaussian pdf $\phi_{\text{DG}}(c)$ (shaded).}
\end{figure*}

For an inhomogeneous Poisson process with rate $r(t)$ conditioned on a common input $c(t)$, the probability of at least one spike occurring in the interval $[t,t+\Delta t]$ is:
\begin{align}
P(\text{spike}\in\Delta t | c ) &= 1 - \exp{\left(-\int_t^{t + \Delta t} \! r(s) \dd s \right)}\\
&= 1 - \exp (-\mathcal{S}) \equiv \tilde{L}(\mathcal{S}),\label{lfunc}
\end{align}
where we have defined: $\mathcal{S} = \int_t^{t + \Delta t} \! r(s) \dd s $.  

Conditioned on the common input -- or, equivalently, the windowed firing rate $\mathcal{S}$ -- each of the $N$ neurons produces spikes independently. Thus, the probability of $k$ cells firing simultaneously is:
\begin{equation}
P_{\text{LNL}}(k) = \binom{N}{k} \int_{-\infty}^{~\infty} \phi_{LNL}(\mathcal{S}) \big(1-\tilde{L}(\mathcal{S})\big)^{n-k} \tilde{L}(\mathcal{S})^{k} \dd \mathcal{S}, \label{kspikes}
\end{equation}
where $\phi_{LNL}(\mathcal{S})$ is the probability density function for $\mathcal{S}$.  We estimate $\phi_{LNL}$ numerically.


Figure~\ref{fig:eiffilter}(a) shows that the LNL cascade captures the general structure of the EIF population output across a range of population sizes.  In particular, it produces an order-of-magnitude improvement over the PME model -- see JS-divergence values in Fig.~\ref{fig:eiffilter}(b) -- and reproduces the skewed structure produced by beyond-pairwise correlations.  

This said, the LNL model does not produce a perfect fit to the EIF outputs, the most obvious problem being the overestimation of the zero spike probabilities, which in the $N=100$ case are overestimated by almost $100\%$ (the tail probabilities are also underestimated). Notably, the LNL fits become almost perfect for lower correlations i.e.~$\rho = 0.05$ (data not shown). This suggests the discrepancies are due to the way the static non-linearity deals with fluctuations to very low or high firing rates $r(t)$; these fluctuations are smaller at lower correlation values, which lead to smaller signal currents in the LNL formulation.  

\medskip 

\paragraph*{The spiking neuron and the Dichotomous Gaussian models produce closely matched population activity.}


So far we have studied the emergence of higher-order correlations in two spiking neuron models -- the EIF model, described in terms of a stochastic differential equation, and the LNL model, which is a continuous-time reduction of the EIF to a doubly-stochastic point process.  Next, we show how these results connect to earlier findings for a more general and abstracted statistical model.  This is the Dichotomous Gaussian (DG) model, which has been shown analytically to produce higher-order correlations and to describe empirical data from neural populations~\cite{Amarietal03,macke11,Yu:2011ii,BarreiroGRS10}.  

In the DG framework, spikes either occur or fail to occur independently and discretely in each time bin.  Specifically, at each time $N$ neurons receive a correlated Gaussian input variable with mean $\gamma$ and correlation $\lambda$. Each neuron applies a step nonlinearity to its inputs, spiking only if its input is positive. Input parameters $\gamma$ and $\lambda$ are chosen to match two target firing statistics:  the spike rate $\mu$ and the the correlation coefficient is $\rho$.

In Fig.~\ref{fig:lnldgcomp}(a), we compare the population output of the DG model with that from the EIF model.  We see that, once the two models are constrained to have the same pairwise correlation $\rho$ and firing rate $\mu$, the rest of their population statistics match almost exactly.  Fig.~\ref{fig:lnldgcomp}(b) provides another view into the similar population statistics produced by the different models.  Here, we study how the heat capacity (i.e., $C = \text{Var}\big(\log_2 P_\beta (x)\big) / n$, at some temperature $T=1/\beta$) varies with population size in Fig.~\ref{fig:lnldgcomp}.  In prior work \cite{macke11} showed that this statistic grows linearly (i.e., extensively) with population size for the DG model, and the Figure shows that the same holds for the EIF and LIF models.  This growth stands, as first noted by \cite{macke11}, in marked contrast to the heat capacity for the Ising model, which saturates at a population of approximately $N = 30$ neurons.

We next develop the mathematical connection between the DG and spiking neuron models, via our description of the LNL model above.  

First, we note that, for the DG model, the correlated Gaussian can be written as: $Z_i = \gamma + \sqrt{1-\lambda} T_i + \sqrt{\lambda} c$ where $T_i$ is the independent input and $c$ is the common input. The probability of a spike is given by $P(Z_i > 0 | c)$ and again we can define a ``$L$" function similar to that in Eqn.~\eqref{lfunc}: 
\begin{equation}
L(c) = P\left( T_i > \frac{-c-\gamma}{\sqrt{1-\lambda}} \right) = \text{CDF}\left(\frac{c+\gamma}{\sqrt{1-\lambda}}\right) \;.
\label{e.Ldg}
\end{equation}
Equipped with Eqn.~\eqref{e.Ldg}, the probability of observing a spike count $k$ is the similar to Eqn.~\eqref{kspikes}:
\begin{equation}
P_{\text{DG}}(k) = \binom{N}{k} \int_{-\infty}^{~\infty} \phi_{DG}(c) \big(1-{L}(c)\big)^{n-k} {L}(c)^{k} \dd c, \label{kspikes2}
\end{equation}
where $\phi_{\text{DG}}(c)$ is the pdf of a one-dimensional Gaussian with mean $0$ and variance $\lambda$. 

We next compare the population spike count distributions $P_{\text{LNL}}(k)$ and $P_{\text{DG}}(k)$. To make the comparison we must transform from the probability density function of the linear-nonlinear model $\phi_{LNL}$ to the Gaussian pdf $\phi_{DG}$ using the nonlinear change of variable:
\begin{equation}
\mathcal{S} = f(c),\quad\text{where}\quad f^\prime(c) = \frac{\phi_{DG}(c)}{\phi_{LNL}\big(f(c)\big)}.
\end{equation}
Writing the LNL cascade probability in terms of the $c$ variable we obtain:
\begin{equation}
P_{\text{LNL}}(k) = \binom{N}{k}\!\!\int_{-\infty}^{~\infty} \phi_{\text{DG}}(c) \big(1-\tilde{L}(f(c))\big)^{n-k} \tilde{L}(f(c))^{k} \dd c 
\end{equation}

Thus, after the transformation the only difference between the LNL and DG models is the functions $L(c)$ vs. $\tilde L( f(c))$.  These functions largely agree over about $2$ standard deviations of the Gaussian pdf of values of the common input signal $c$~\footnote{At large values of common input $c$, the higher values of $L(c)$ for the DG model account for its closer correspondence to the EIF model in the tails of $P(k)$ in Figure~\ref{fig:lnldgcomp}(a).}.  

This reveals why the LNL and DG -- and, by extension, the LIF -- models all produce such similar population-level outputs, including their higher-order structure.

The success of the DG model in capturing EIF population statistics is significant for two reasons.  First, it suggests why this abstracted model has been able to capture the population output recorded from spiking neurons.  Second, because the DG model is a special case of a Bernoulli generalized linear model (see appendix), our finding indicates that this very broad and easily fittable class of Bernoulli statistical models may be able to capture the higher-order population activity in neural data.

\paragraph*{Summary and conclusion.}

We have shown that Exponential-Integrate and Fire (EIF) neurons receiving common input give rise to strong higher-order correlations.  Moreover, the correlation structure that results can be predicted from a linear-nonlinear cascade model, which forms  a tractable reduction of the EIF neuron.  Beyond giving an explicit formula for the EIF population spike-count distribution, our findings for the cascade model demonstrate that common input will drive higher-order correlations in a widely-used class of point process models.  Finally, we show that there is a surprisingly exact connection between the population dynamics of the EIF and cascade models and that of the (apparently) simpler Dichotomized Gaussian (DG) model of ~\cite{macke11,Amarietal03}, which has  been successful in explaining data from neural recordings.  Taken together, these findings are encouraging for the goal of connecting circuit mechanisms and the statistics of the activity that they produce.
   
\smallskip 

The authors thank Liam Paninski for helpful insights. This work was supported by the Burroughs Wellcome Fund Scientific Interfaces Program and the NSF grant CAREER DMS-1056125.  

\bibliographystyle{apalike}
\bibliography{HOC_bib}


\noindent \appendix{{\it Appendix: Generalized Linear models.}}  The LNL model provides a reduction of the EIF model to an inhomogeneous poisson process that is based directly on the underlying SDEs, and is in extremely wide use in neural modeling~\cite{Day+01}.  However, it far from the only approach to statistical modeling of spiking neurons.  In particular, generalized linear models can be fit to the Bernoulli data given by the 1's and 0's of binned spikes in individual cells.  Such models similarly apply a linear filter to the common input signal, and followed by a static nonlinearity $f(\cdot)$, to yield a spiking probability for the current time bin.  Noting that any linear filter on our (gaussian white noise) input signal will yield a gaussian value $s$, this class of models therefore yields spiking probabilities $f(s)$ where $s$ is gaussian.  Comparing with Eqn.~\eqref{e.Ldg} in the main text, we see that the DG and generalized linear models have the same general form, when $f$ is taken to be the cumulative distribution function for a gaussian (as in ``probit" models).

\end{document}